\documentstyle[epsf]{mn}

\def\figdir{.}

\def\na#1{{#1}}

\def\dim#1{\mbox{\,#1}}

\def\epsscale#1{\epsfxsize=#1\columnwidth}
\def\plotone#1{\par\centerline{\epsfbox{#1}}}

\title[Power Spectrum from the Ly-$\alpha$ Forest]{Matter Power Spectrum from the Lyman-Alpha Forest: Myth or Reality?} 
\author[Gnedin \& Hamilton]{Nickolay Y.\ Gnedin$^1$ and Andrew J.\ S.\
Hamilton$^2$\\
$^1$Center for Astrophysics and Space Astronomy, 
University of Colorado, Boulder, CO 80309; {\tt gnedin@casa.colorado.edu}\\
$^2$JILA, University of Colorado, Boulder, CO 80309; 
{\tt Andrew.Hamilton@colorado.edu}}

\begin{document}

\maketitle

\begin{abstract}
We investigate possible systematic errors in the recent measurement
of the matter power spectrum from the Lyman-alpha forest by
Croft et al.\ \shortcite{CWB01}.
We find that for a large set of prior cosmological
models the Croft et al.\ result holds quite well, with systematic errors
being comparable to random ones, when a dependence of the recovered
matter power spectrum on the cosmological parameters at $z\sim3$
%(missed by Croft et al.)
is taken into account. We find that 
peculiar velocities cause the flux power spectrum to be smoothed
over about $100-300\dim{km/s}$, depending on scale. 
Consequently, the recovered matter
power spectrum is a smoothed version of the underlying true power spectrum.
Uncertainties in the recovered power spectrum are thus correlated over
about $100-300\dim{km/s}$.

As a side effect, we find that residual fluctuations in the ionizing
background, while having almost no effect on the recovered matter
power spectrum, significantly bias estimates of the baryon density
from the Lyman-alpha forest data.

We therefore conclude that the Croft et al.\ result provides a powerful
new constraint on cosmological parameters and models of structure
formation.
\end{abstract}


\section{Introduction}

The Lyman-alpha forest is perceived as being simple, and in this
simplicity promises to become a key to many unsolved problems in cosmology.
In the most recent demonstration of the power of simplicity, 
Croft et al.\ \shortcite{CWB01} 
presented a measurement of the matter linear power
spectrum on scales which are nonlinear today -- a measurement not easily
reproducible by other means, such as galaxy surveys at low redshift.

Not only was the phenomenon that they modeled simple, but also the method
that they used to recover the linear power spectrum was simple
-- perhaps excessively simple.
For example, Croft et al.\ \shortcite{CWB01} considered only one prior
cosmological model in deriving the correction function that translates the
observed flux power spectrum into a linear power spectrum. 

In this paper our purpose is to check whether their method is reliable 
by sampling over a range of prior cosmological models. We also attempt to
estimate systematic errors and their covariance. We begin in \S 2 by
making sure that we are able to reproduce Croft et al.\ \shortcite{CWB01} results.
In \S 3 we give our results. We close in \S 4 with an optimistic 
conclusion.

\section{Method}
\label{sec:met}

Since our goal is to evaluate the accuracy of the Croft et al.\ \shortcite{CWB01} method,
we apply their method to a range of cosmological models.
We adopt the same linear transfer function,
but we vary the amplitude and the tilt,
and we experiment with different geometries and different Hubble constants.
In \S\ref{bandpower} we also report the effect of varying the linear power
spectrum over narrow intervals of wavenumber.
We use a standard Particle-Mesh method to simulate the distribution the
matter at $z=2.72$. Following Croft et al.\ \shortcite{CWB01}
we assume that the underlying baryon density follows that of the dark matter.
At small scales the baryon density is smoothed out compared to the
dark matter density, but the scale at which this occurs,
the ``filtering'' scale, is much smaller than the range of scales
of interest here \cite{GH98} 
(we discuss this in detail in \S\ \ref{sec:inh}).
All our simulations have a box size of $20h^{-1}\dim{Mpc}$ (in comoving 
coordinates) and $256^3$ particles on a mesh of the same size.
The gravitational force is calculated with a Green function
that includes the Optimal Antialising filter \cite{FB94},
by summing over three Brilloun zones.

\na{
In calculating the synthetic Lyman-alpha absorption spectra we follow
the the Croft et al.\ \shortcite{CWB01} procedure in precise detail: we
compute the synthetic spectra for 1000 lines of sight which cross the
computational box at random angles (to avoid bias introduced for lines of
sight parallel to the box sides) for an assumed value for the spatial
homogeneous photoionization rate $\Gamma$. We then compute the mean opacity 
$\bar\tau(\Gamma)$ and find the value of $\Gamma$ that gives the desired value
for the mean opacity ($\bar\tau=0.349$ except as described in \S
\ref{sec:meanop}). We then compute the 1D flux power spectra of the
normalized flux for each
line of sight using a Lomb periodogram and average the results over 1000
lines of sight. We have verified that this number -- 1000 lines of sight --
is fully sufficient to obtain numerical results with the precision well
below the observational error.
}

Before we can test for systematic errors in the Croft et al.\ \shortcite{CWB01} method,
we must first check that we can reproduce their results accurately. 

In order to reduce the uncertainty in the mean flux power spectrum, an
average of several different random realizations is usually used. Normally,
this procedure requires a large number of realizations, because the 
variance decreases only as the square root of the number of realizations.
Instead, we adopt a slightly different approach. Out of all random
realizations, we first choose ``good'' ones, i.e.\ those that give
a measured linear power spectrum close to the
input linear power spectrum. We choose 3 ``best'' realizations 
out of 100 random ones, and an average over those three recovers
the input power spectrum at least as well as a plain average over 100
truly random realizations.

\begin{figure}
\epsscale{1.0}
\plotone{\figdir/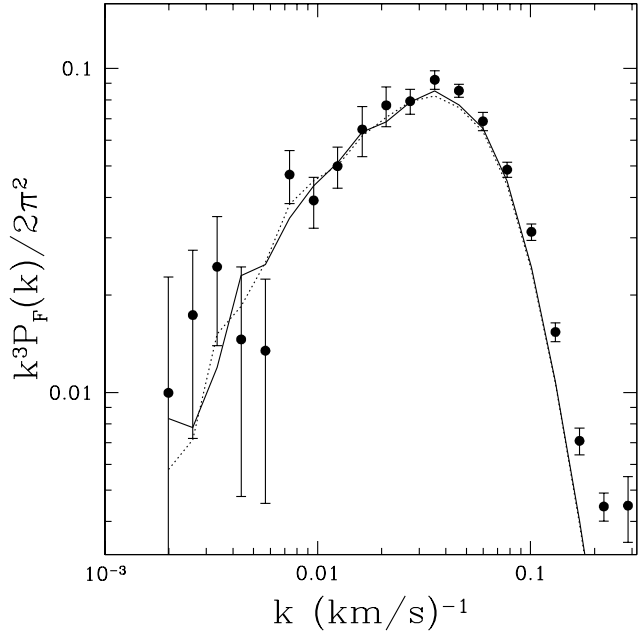}
\caption{\label{figRL}
The flux power spectrum predicted by the Croft et al.\ \shortcite{CWB01} fiducial model,
compared to the observed flux power spectrum from Croft et al.\ \shortcite{CWB01}.
The {\em dotted line\/} is the model averaged over 3 best realizations
(roughly equivalent to 100 random realizations)
while the {\em solid line\/} is averaged over 12 best realizations
(roughly equivalent to 400 random realizations).}
\end{figure}
Figure \ref{figRL} shows the flux power spectrum $P_F(k)$
(shown as $\Delta^2(k)\equiv k^3P(k)/2\pi^2$)
averaged over 3 and over 12 best realizations (equivalent of
100 and 400 random realizations). The difference between the two curves
provides an estimate of about 1/2 of the uncertainty in the mean flux 
power spectrum 
(because averaging over 12 realizations gives half the variance
of averaging over just 3 realizations). We notice that this difference is
significantly smaller than the statistical error-bars of the observational
data, indicating that our flux power spectrum is calculated
with sufficient precision.

Hereafter we use three ``good'' realizations per normalization for each 
cosmological model. Thus, each model requires at least six simulations
(two different normalizations above and below the best-fit value), and more
if our initial choice for the two normalizations does not bracket the
best-fit value.

\begin{figure}
\epsscale{1.0}
\plotone{\figdir/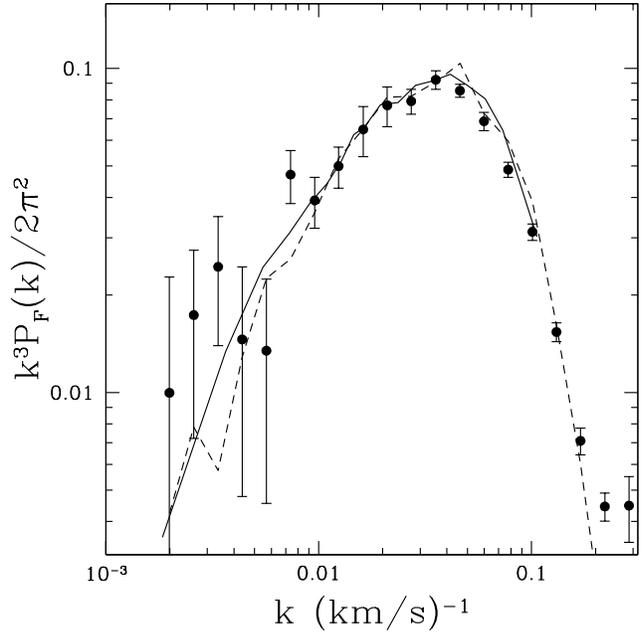}
\caption{\label{figRF}The flux power spectrum from the Croft et al.\
\shortcite{CWB01} fiducial model ({\it solid line\/}) and our reproduction of
their result ({\it dashed line\/}).
}
\end{figure}
As evidence that our method reproduces the Croft et al.\ \shortcite{CWB01}
results, we show in Figure \ref{figRF} their fiducial model 
(courtesy Rupert Croft) and the flux power spectrum from our simulation
of exactly the same cosmological model. Notice that there exist a 
difference on large scales, which is most likely due to the 
smaller size of our
computational box, but that it is smaller than the observational
errors and so is unimportant.

\section{Results}

\subsection{Systematic errors of the recovered linear power spectrum}

In this paper our main concern
is the effect that an assumption of a specific cosmological model makes
on recovering the linear power spectrum from the flux power spectrum.
Specifically, Croft et al.\ \shortcite{CWB01} assume that
the flux power spectrum $P_F(k)$ is proportional to the linear
power spectrum $P_L(k)$ at the same wavenumber
\begin{equation}
	P_F(k) = b^2(k)P_L(k)
	\label{pfl}
\end{equation}
the bias factor $b(k)$ being independent of, or at least insensitive to,
the power spectrum $P_L(k)$.

If the ``bias factor'' $b(k)$ were independent of the linear power
spectrum, this would be a direct analogy to the biased galaxy linear
power spectrum. However, {\it this is not the case\/} - the ``bias factor''
$b(k)$ depends on the amplitude of the linear power spectrum $P_L(k)$,
so that the relation (\ref{pfl}) is {\it not\/} a linear relation.
Correctly, equation (\ref{pfl}) should be written as
\begin{equation}
	P_F(k) = b^2[k,P_L]P_L(k),
	\label{pflc}
\end{equation}
where we use the square brackets to underline that $b$ is actually an 
{\it operator\/} acting on the linear power spectrum $P_L(k)$. Thus, the
recovered ``observed'' linear power spectrum,
\begin{equation}
	P_L^{\rm obs}(k) = {P_F^{\rm obs}(k)\over b^2[k,P_L]},
	\label{howwedoit}
\end{equation}
actually depends on the assumed linear power spectrum, and this dependence
appears as a systematic error which is not included in the estimate of
the error bars by Croft et al.\ \shortcite{CWB01}.

For a particular choice of $P_L$ that fits the observed flux power 
spectrum, relationship (\ref{pflc}) can be linearized. 
For small deviations $\Delta P_L$
about an assumed linear power spectrum $P_L$ we obtain
\begin{equation}
	\Delta P_F(k) = \sum_{k^\prime} b^2(k,k^\prime)\Delta P_L(k^\prime),
	\label{pfll}
\end{equation}
where the matrix $b(k,k^\prime)$ can be obtained by differentiating
equation (\ref{pflc}). Even in this linearized form this equation is
more complicated that the Croft et al.\ \shortcite{CWB01} assumption (\ref{pfl}).

\na{
Throughout this paper, when we refer to the ``recovered linear power
spectrum'', we always mean $P_L^{\rm obs}(k)$ obtained via equation
(\ref{howwedoit}). The sole purpose of this paper is to investigate how
the dependence of $b$ on $P_L$, ignored by Croft et al.\ \shortcite{CWB01},
affects the recovery procedure.
}

\subsubsection{Dependence on the Hubble constant or spectral curvature}

\begin{figure}
\epsscale{1.0}
\plotone{\figdir/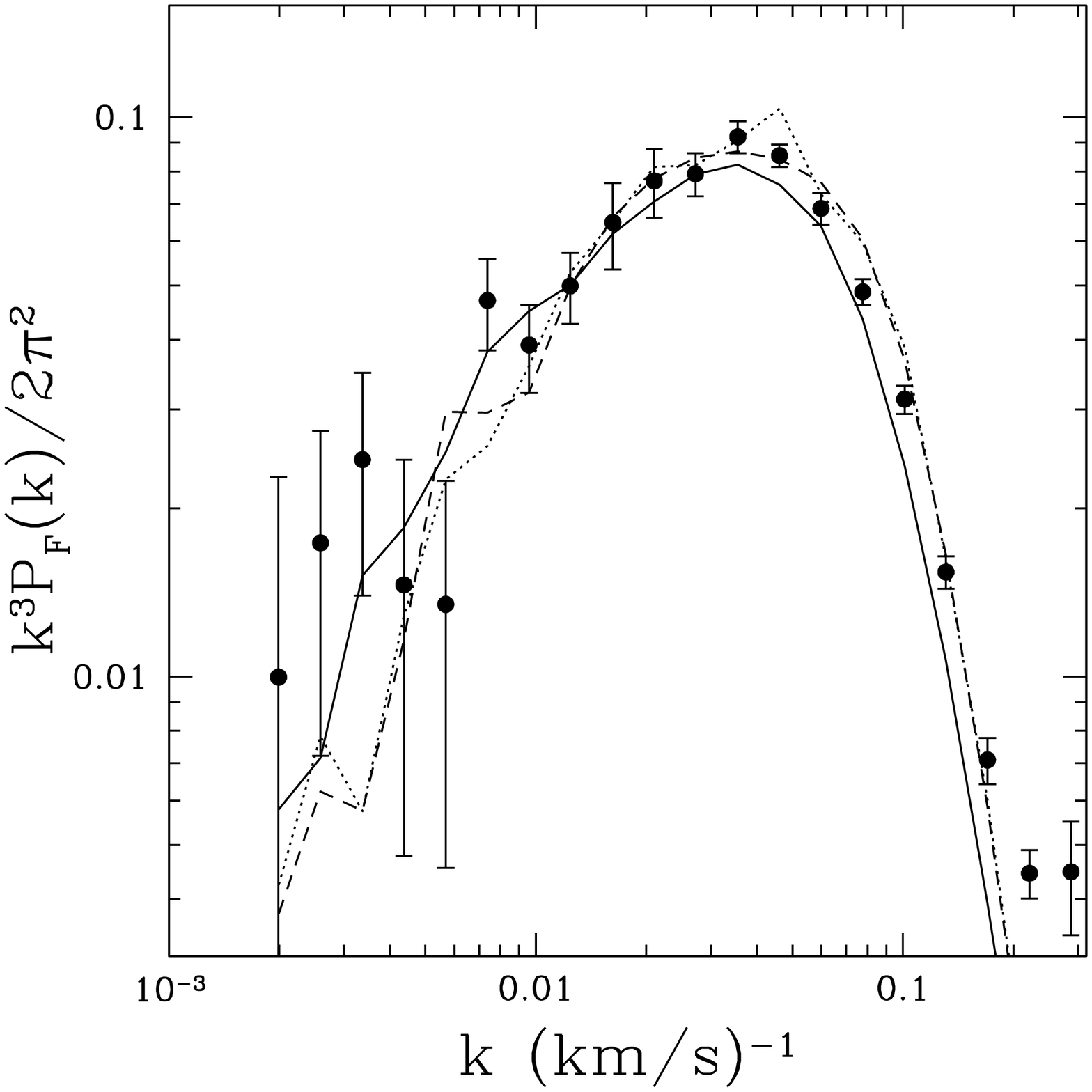}
\epsscale{1.0}
\plotone{\figdir/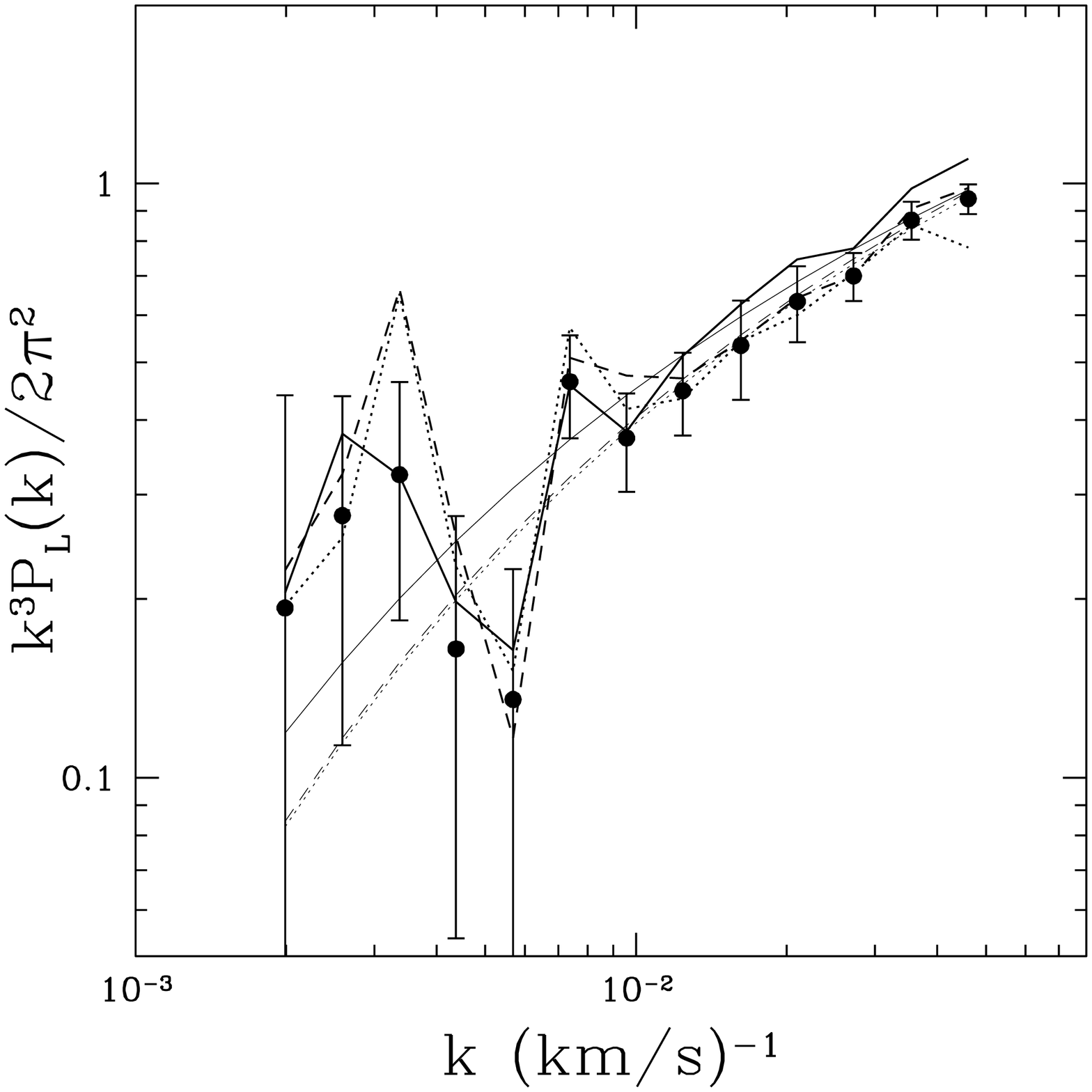}
\caption{\label{figHF}({\it a\/}) The flux power spectrum and ({\it b\/}) the
recovered linear power spectrum $P_L^{\rm obs}$ in 
the Croft et al.\ \shortcite{CWB01} fiducial model (EdS cosmology; {\it solid
line\/}),
the same cosmological model with the Hubble constant rescaled to a value
appropriate to a flat cosmology with 
$\Omega_{m,0}=0.4$ (just like in Croft et al.; {\it dotted line\/})),
and for the flat cosmological model with $\Omega_{m,0}=0.4$
({\it dashed line\/}). Thin solid lines in this and all following
figures show the assumed linear power spectrum $P_L$ for the underlying
cosmological model. 
}
\end{figure}
An important part of the Croft et al.\ \shortcite{CWB01} procedure is the translation
from the spatial units used in a simulation to velocity units of the
observational data. This requires adopting a value for the Hubble constant
at $z=2.72$ 
\na{
which is used to scale spatial scales to velocity space. Specifically, 
following Croft et al.\ \shortcite{CWB01}, we adopt
\begin{equation}
    v_{{\rm tot},i} = H_3\left(r_i + v_{{\rm pec},i}/H_{{\rm EdS},3}\right),
\end{equation}
where $r_i$ is the physical position of a pixel $i$ in the synthetic
spectrum, $v_{{\rm tot},i}$ is the velocity position of the same pixel,
$v_{{\rm pec},i}$ is the peculiar velocity of a given pixel in the
simulation (in the EdS cosmology), $H_3$ is the adopted value of the Hubble
constant at $z=2.72$, and $H_{{\rm EdS},3}=720h\dim{km/s/Mpc}$ is the Hubble
constant in the EdS cosmology at $z=2.72$. This procedure is identical to
the one adopted in Croft et al.\ \shortcite{CWB01}: we have applied our
procedure to the simulation data kindly provided to us by Rupert Croft, and
we obtained numerically indistinguishable results.
}

\na{
It is important to emphasize here that this rescaling of the Hubble
constant only affects the transformation from position to velocity
space. The primordial power spectrum of fluctuations as a function of
position adopted in the simulations is not changed, but rescaling of the
velocity units does result in the linear power spectra being different in
velocity space.}

It can be shown that for pure power-law power
spectra all the dependence on the Hubble constant cancels out exactly.
However, this is not true for a general power spectrum with 
varying slope. In order to test this dependence we have run a
low density flat cosmological model in addition to the fiducial
model. Both models have identical initial conditions, and both run in
the matter-dominated universe ($\Omega_{m,3}=1$ for both models). They
differ only by the values of the Hubble constant
$H=H_0\sqrt{\Omega_{m,0}}(1+z)^{3/2}$, which are different by a factor
$\sqrt{0.4}=0.6$. Croft et al.\ \shortcite{CWB01} did not use a 
Hubble constant appropriate to their cosmological model.
Rather, they ran the simulation in the EdS cosmology, but
adopted the Hubble constant from a flat cosmology with $\Omega_{m,0}=0.4$.
We also show the so rescaled model in Fig.\ \ref{figHF}
(which is the model also shown in Fig.\ \ref{figRF}).
As can be seen from Figure \ref{figHF}, the three
recovered power spectra differ by less than the random error of the
measurement, and we thus conclude that the dependence on the Hubble
constant (or, equivalently, on the rate of change in the local slope - but
see \S\ \ref{bandpower}) is insignificant.

\subsubsection{Assumed effective equation of state}

\begin{figure}
\epsscale{1.0}
\plotone{\figdir/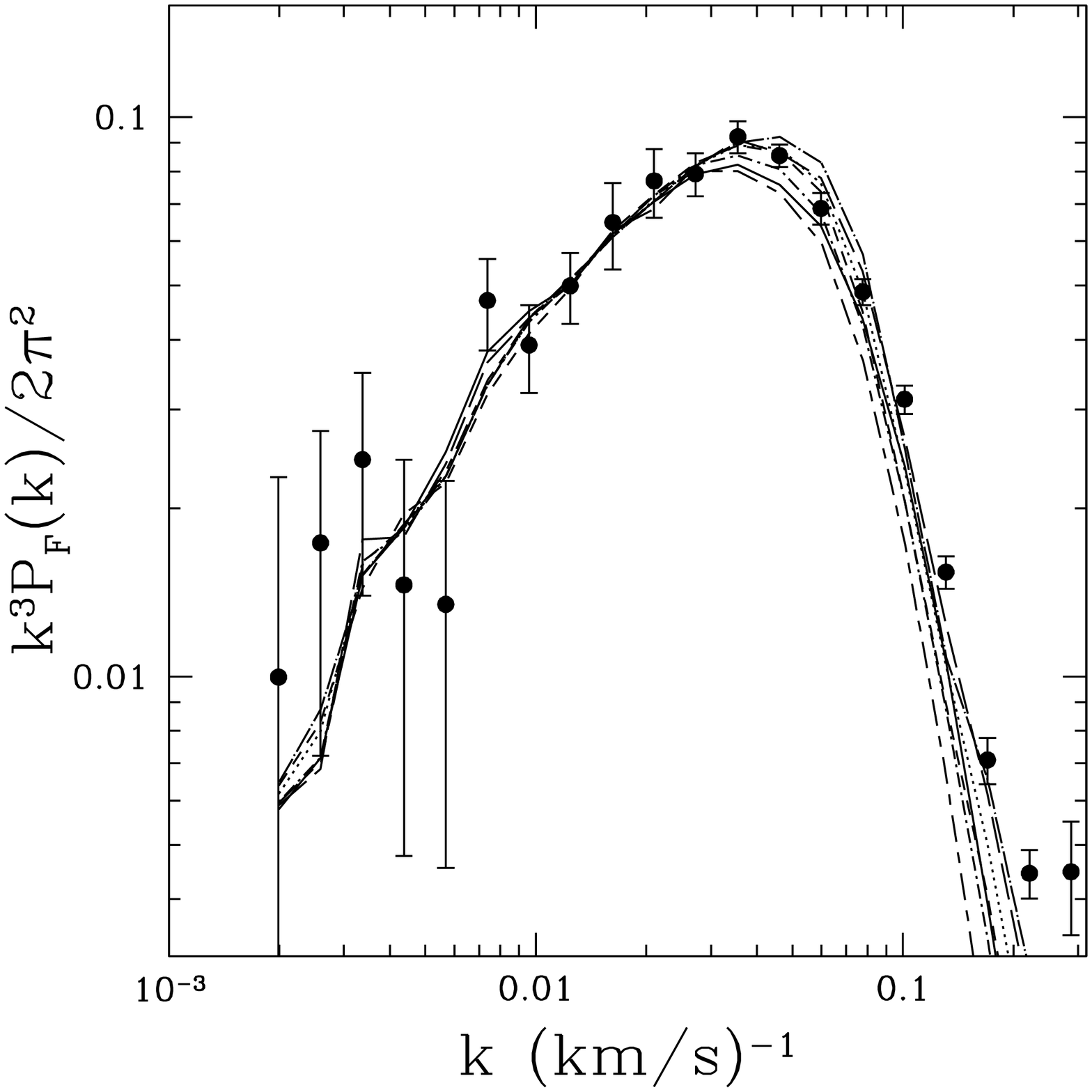}
\epsscale{1.0}
\plotone{\figdir/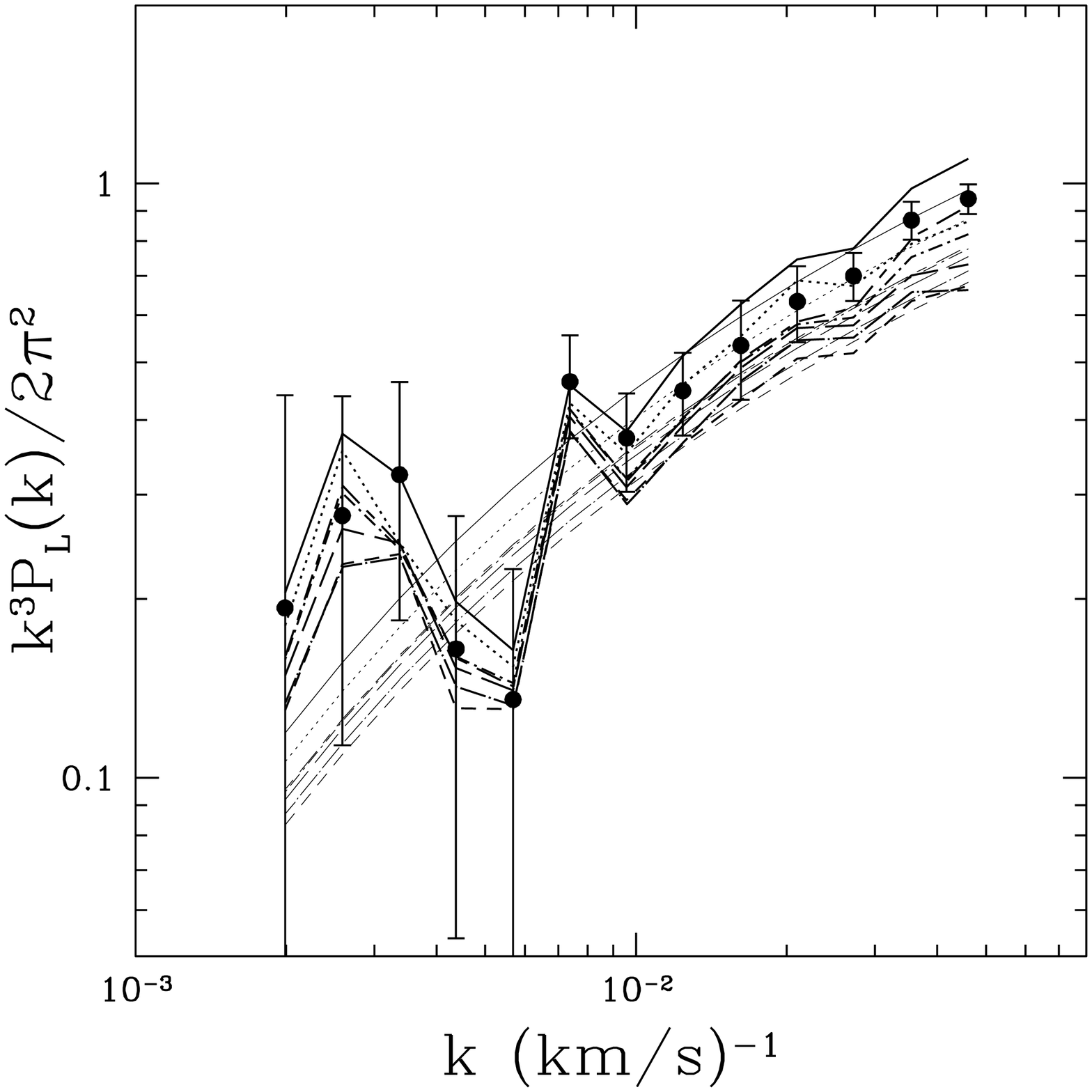}
\caption{\label{figEF}The flux power spectrum ({\it a\/}) and the
recovered linear power spectrum ({\it b\/}) in 
the Croft et al.\ \shortcite{CWB01} model with different assumed
effective equations of state:
$T_0=15{,}000\dim{K}$, $\gamma=1.6$ ({\it solid line\/}, Croft et al.\ assumed values),
$T_0=20{,}000\dim{K}$, $\gamma=1.2$ ({\it dotted line\/}), 
$T_0=23{,}000\dim{K}$, $\gamma=1.2$ ({\it short-dashed line\/}), 
$T_0=17{,}000\dim{K}$, $\gamma=1.2$ ({\it long-dashed line\/}), 
$T_0=20{,}000\dim{K}$, $\gamma=1.4$ ({\it dot -- short-dashed line\/}), and
$T_0=20{,}000\dim{K}$, $\gamma=0.9$ ({\it dot -- long-dashed line\/}).
$T_0=20{,}000\dim{K}$, $\gamma=1.6$ ({\it short-dashed -- long-dashed line\/}).
}
\end{figure}
Croft et al.\ \shortcite{CWB01} point out that uncertainty in the 
``effective equation of state of the IGM'',
i.e.\ in the relationship
between the temperature and the density of the photoionized gas, is one of
the dominant systematic errors in recovering the linear power spectrum.
This
relation is commonly parameterized as \cite{HG98}
\begin{equation}
	T\approx T_0(1+\delta)^{\gamma-1},
\end{equation}
where $T_0$ and $\gamma$ are parameters and $\delta$ is the gas overdensity.
Croft et al.\ \shortcite{CWB01} adopt the values of $T_0=15{,}000\dim{K}$ and
$\gamma=1.6$. However, recent measurements of the effective equation of state
at $z\sim3$ \cite{RGS00,STR00,MMR01} 
suggest somewhat different values for these parameters: 
$T_0=20{,}000\pm3{,}000\dim{K}$ and $\gamma=1.3\pm0.2$ at $z=2.72$
(all three measurements taken together and weighted by their uncertainties). 
We have run the six models
with values for $T_0$ and $\gamma$ in the observed range,
and the results are shown in Figure \ref{figEF}. We find that
the effect of the slope $\gamma$ on the recovered power spectrum is
small, below the random uncertainties of the measurement, and 
so we do not attempt to model it, but regard it as part of 
the total systematic error quoted in Table \ref{tabone}. The amplitude
$T_0$ however does affect the recovered linear power spectrum, which scales
approximately inversely proportionally to $T_0$,
\begin{equation}
	P_L(k;T_0) = P_L(k;20{,}000\dim{K})\left(20{,}000\dim{K}\over 
	T_0\right).
	\label{t0dep}
\end{equation}
Rescaling the recovered power spectra this way reduces the rms difference
between various models to about 1\%.

This dependence is easy
to understand: for a given density and velocity distribution and a given
photoionization rate, higher
temperature means broader absorption lines, which in turn
implies a lower mean opacity. When the mean opacity is renormalized to
the fiducial value 0.349, the photoionization rate is adjusted downward,
and a given value of the flux now corresponds to a lower value of the
density and thus the lower amplitude of the power spectrum (since
readjustment of the photoionization rate scales all densities by the
same factor).

Because the value $T_0=20{,}000\dim{K}$ is favored by observations, we
adopt this value as our fiducial value in the rest of this paper.

\subsubsection{Assumed mean optical depth}
\label{sec:meanop}

\begin{figure}
\epsscale{1.0}
\plotone{\figdir/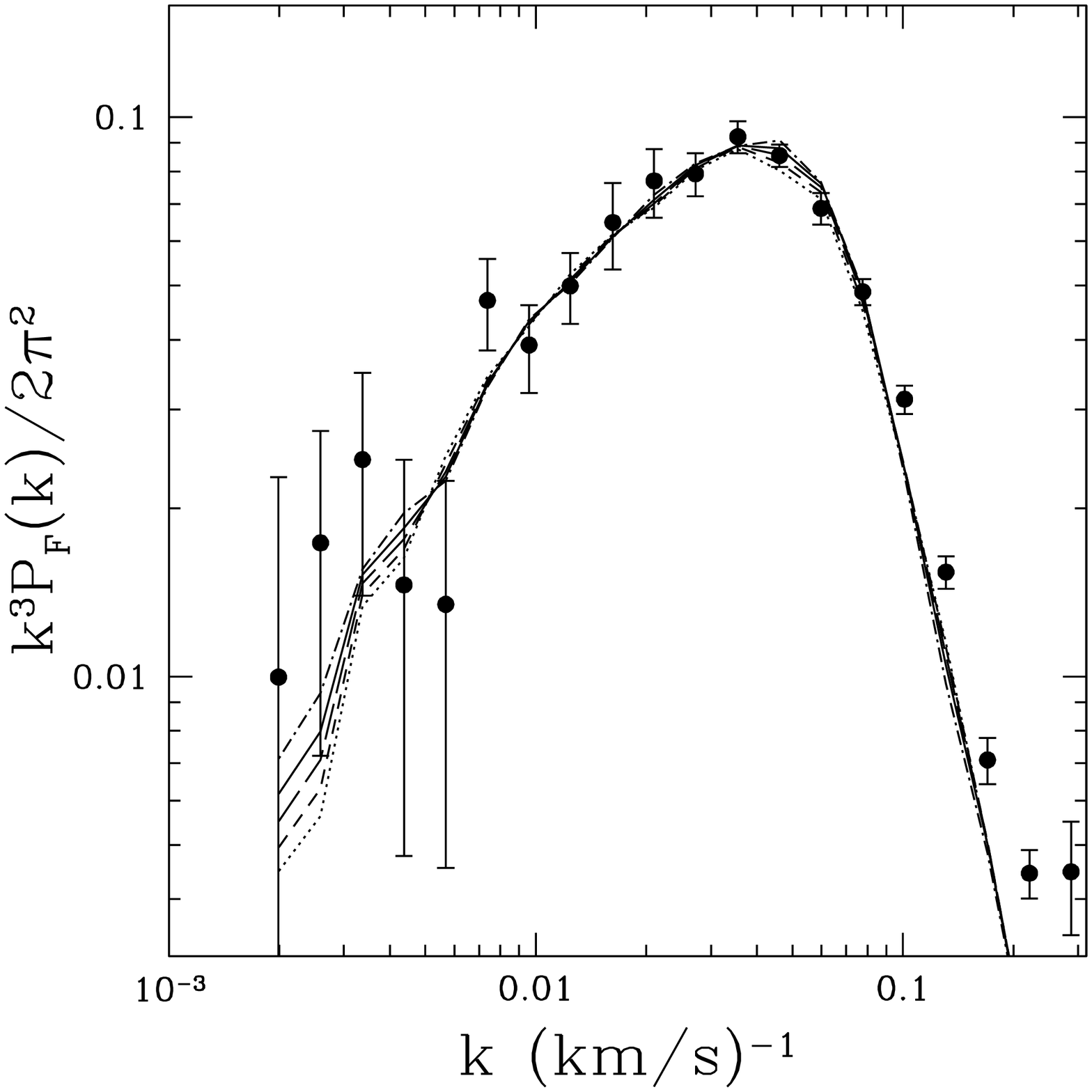}
\epsscale{1.0}
\plotone{\figdir/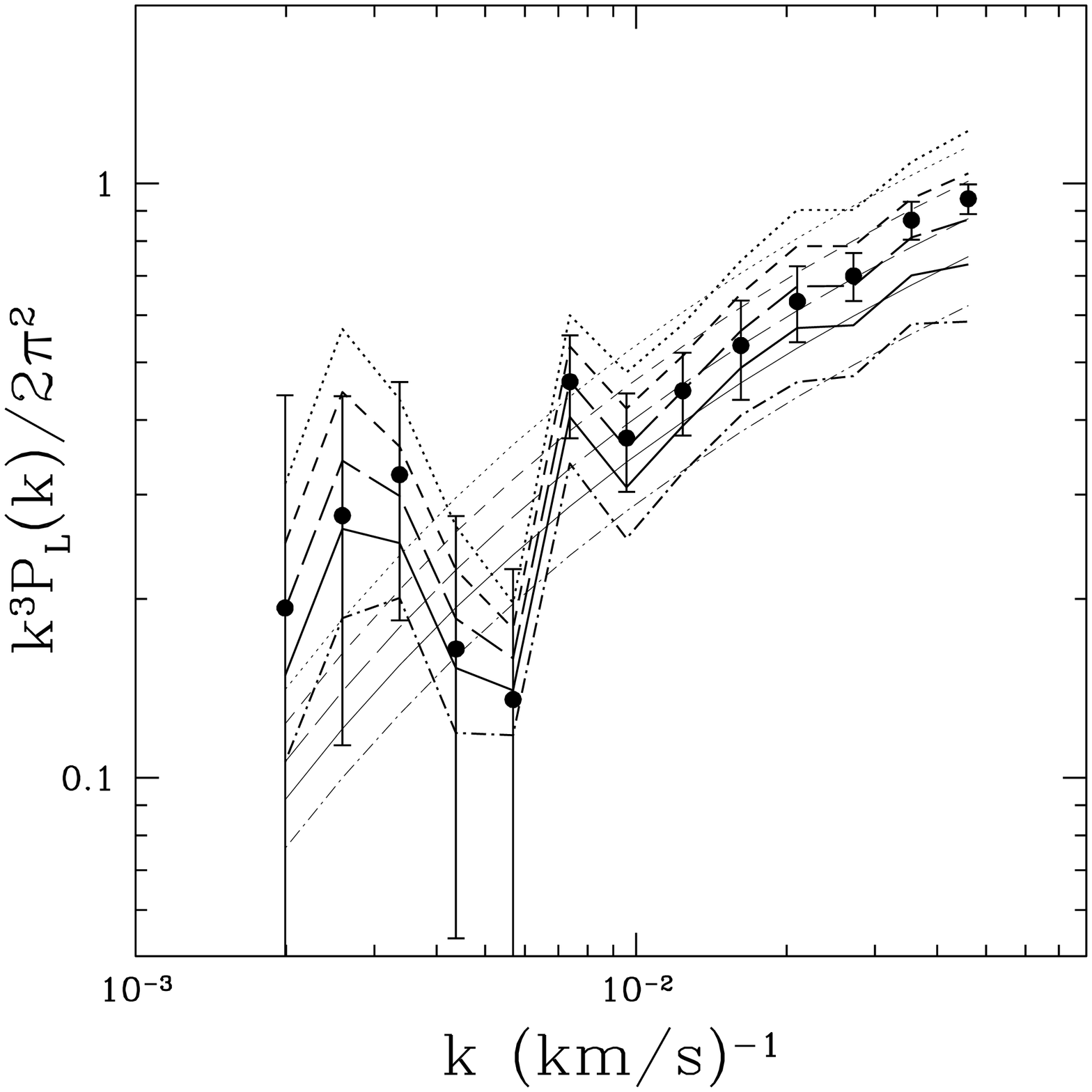}
\caption{\label{figTF}The flux power spectrum ({\it a\/}) and the
recovered linear power spectrum ({\it b\/}) in 
the Croft et al.\ model with different assumed values for
the mean optical depth:
$\tau=0.349$ ({\it solid line\/}, Croft et al.\ assumed value),
$\tau=0.260$ ({\it dotted line\/}), 
$\tau=0.285$ ({\it short-dashed line\/}), 
$\tau=0.315$ ({\it long-dashed line\/}), 
$\tau=0.400$ ({\it dot -- short-dashed line\/}).
}
\end{figure}
Another possible systematic error discussed by Croft et al.\ \shortcite{CWB01} 
is the assumed value for the mean optical depth - the value used to
normalize synthetic absorption spectra. They assumed a value of 
$\tau=0.349$ based on Press, Rybicki, \& Schneider \shortcite{PRS93} 
(hereafter PRS) value. The PRS measurement actually gives 
$\tau_{\rm PRS}=0.349^{+0.051}_{-0.034}$. A more recent measurement by
McDonald et al.\ \cite{MMR00} (hereafter MCD) gives a somewhat lower value,
$\tau_{\rm MCD}=0.285\pm0.025$. In order to test all the parameter range
suggested by observations, we have run four models with the mean optical
depth different from the fiducial value: 
$\tau=0.400$ (PRG+$1\sigma$),
$\tau=0.315$ (PRG-$1\sigma$ and about MCD+$1\sigma$),
$\tau=0.285$ (MCD value), and 
$\tau=0.260$ (MCD-$1\sigma$). The flux power spectra and the recovered
linear power spectra for these four cases and the fiducial case are
given in Figure \ref{figTF}. As one can see, the recovered power
spectra scale with $\tau$. We find a somewhat weaker dependence than
that reported by Croft et al.\ \shortcite{CWB01}, mostly because we only consider a range
of optical depth around the observed values. The following simple power-law
scaling takes out almost all dependence of the recovered linear power spectrum
on $\tau$ (the rms difference
between the rescaled power spectra is less than 1\%):
\begin{equation}
	P_L(k;\tau) = P_L(k;0.349)\left(0.349\over\tau\right)^{0.75}.
	\label{taudep}
\end{equation}

The physical origin of this dependence is also clear: lower $\tau$
means higher densities for the same value of the flux.

\subsubsection{Slope of the prior power spectrum}

\begin{figure}
\epsscale{1.0}
\plotone{\figdir/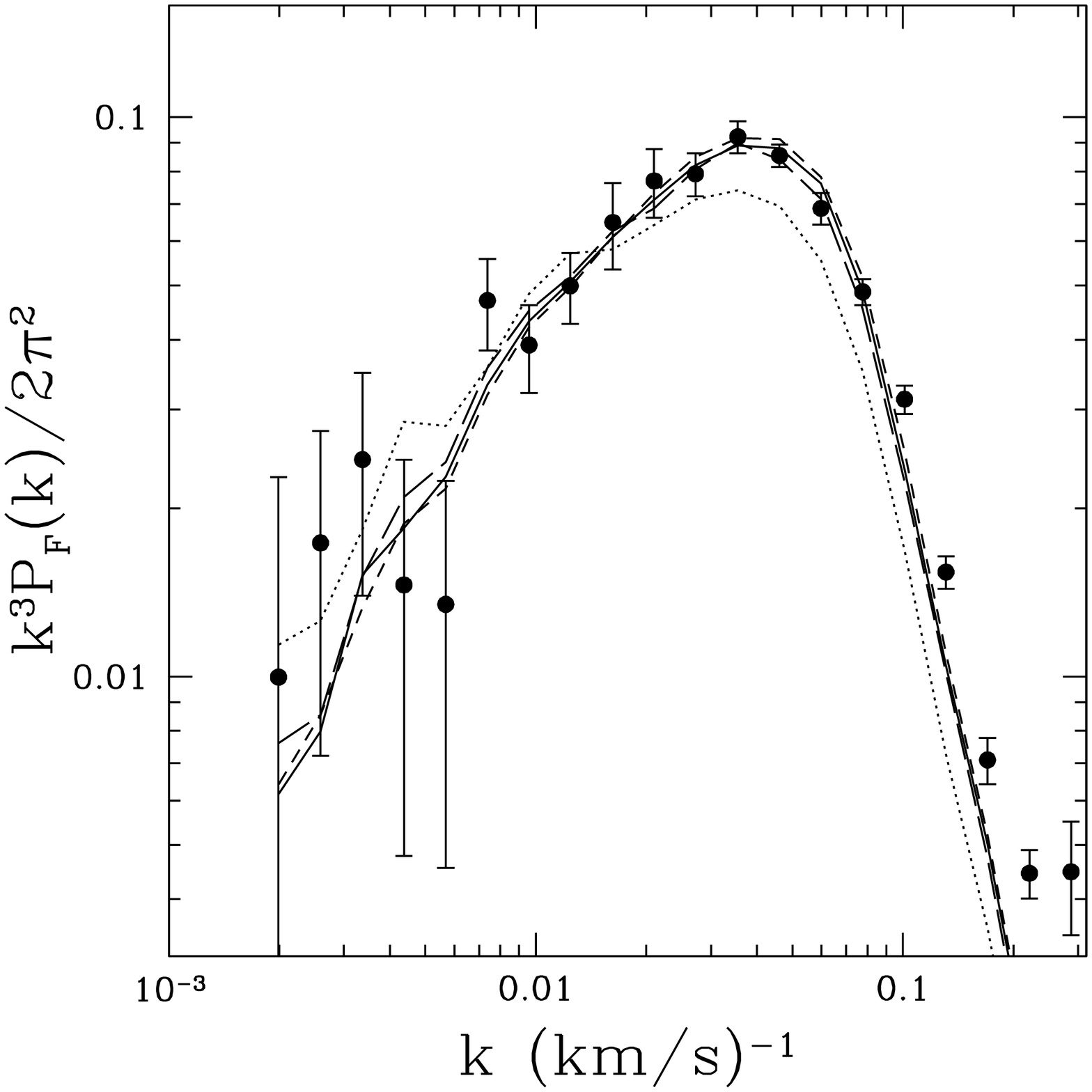}
\epsscale{1.0}
\plotone{\figdir/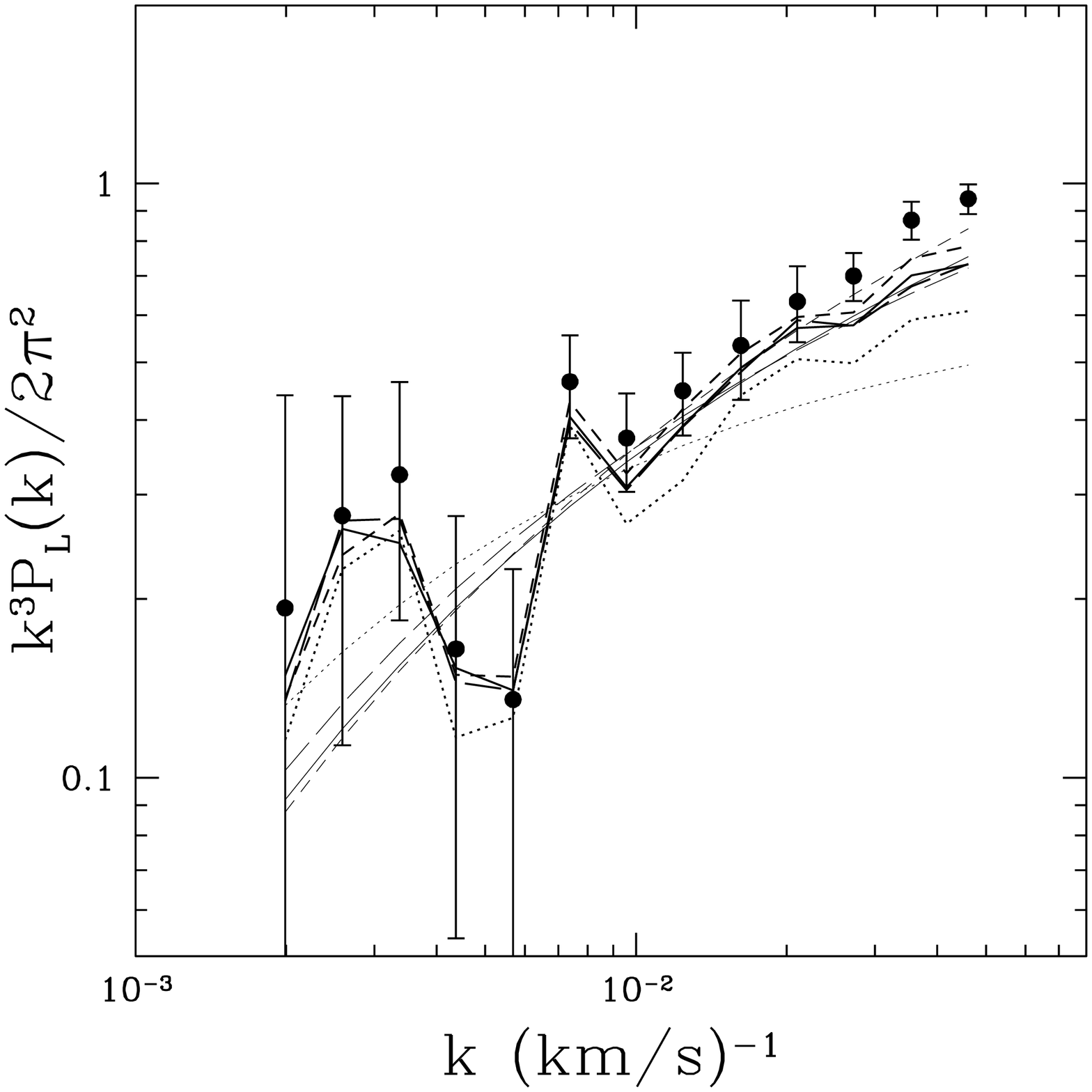}
\caption{\label{figSF}The flux power spectra ({\it a\/}) and the
recovered linear power spectra ({\it b\/}) in four cosmological models
with the different slope of the primordial power spectrum, as shown by
thick lines:
$n=0.95$ ({\it solid line\/}, Croft et al.\ fiducial model),
the same model with 
$n=0.9$ ({\it long-dashed line\/}), 
$n=1.0$ ({\it short-dashed line\/}), and
$n=0.7$ ({\it dotted line\/}). Thin lines in panel ({\it b\/}) show the prior
linear power spectra for these models and symbols show the
data from Croft et al.\ \shortcite{CWB01}.
}
\end{figure}
Another possible source of systematic error is
the slope of the prior linear power spectrum $P_L(k)$. We test
four models, with $n=1.0$, $n=0.95$ (the Croft et al.\ model), $n=0.9$,
and $n=0.7$. Figure \ref{figSF} shows the flux and recovered linear
variance for these models. The $n=0.7$ model does not fit the flux power
spectrum as well as other models, but the fit is still statistically
acceptable ($\chi^2=11$ with 11 degrees of freedom).
In all cases however, the recovered linear power spectrum agrees well
with the fiducial model, indicating that
the method is insensitive to quite significant variations in the
slope of the prior power spectrum.
In other words, the effective bias factor $b(k)$ in equation (\ref{pfl})
appears similar for all the model spectra, an encouraging result.

\subsubsection{Dependence on the matter density}

\begin{figure}
\epsscale{1.0}
\plotone{\figdir/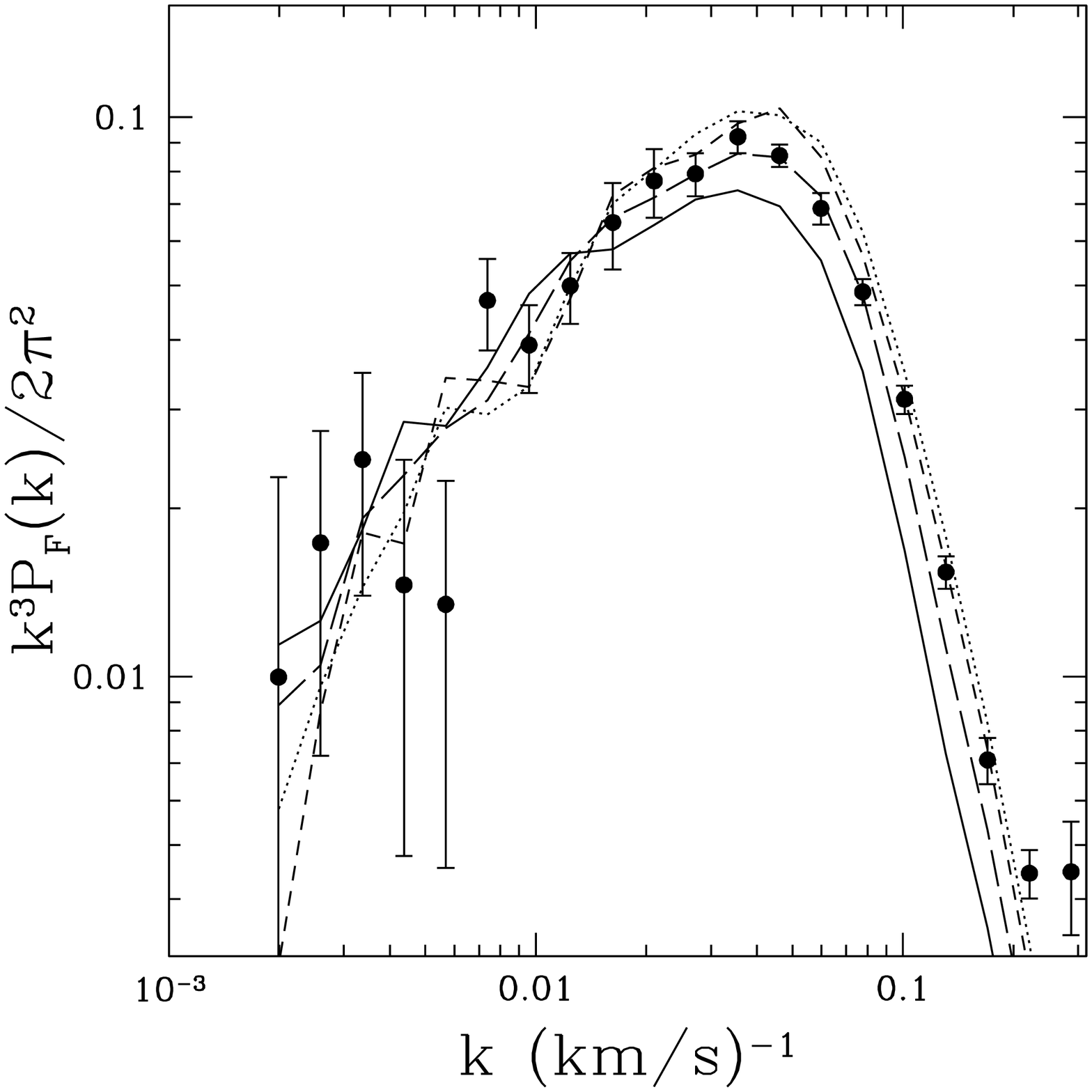}
\epsscale{1.0}
\plotone{\figdir/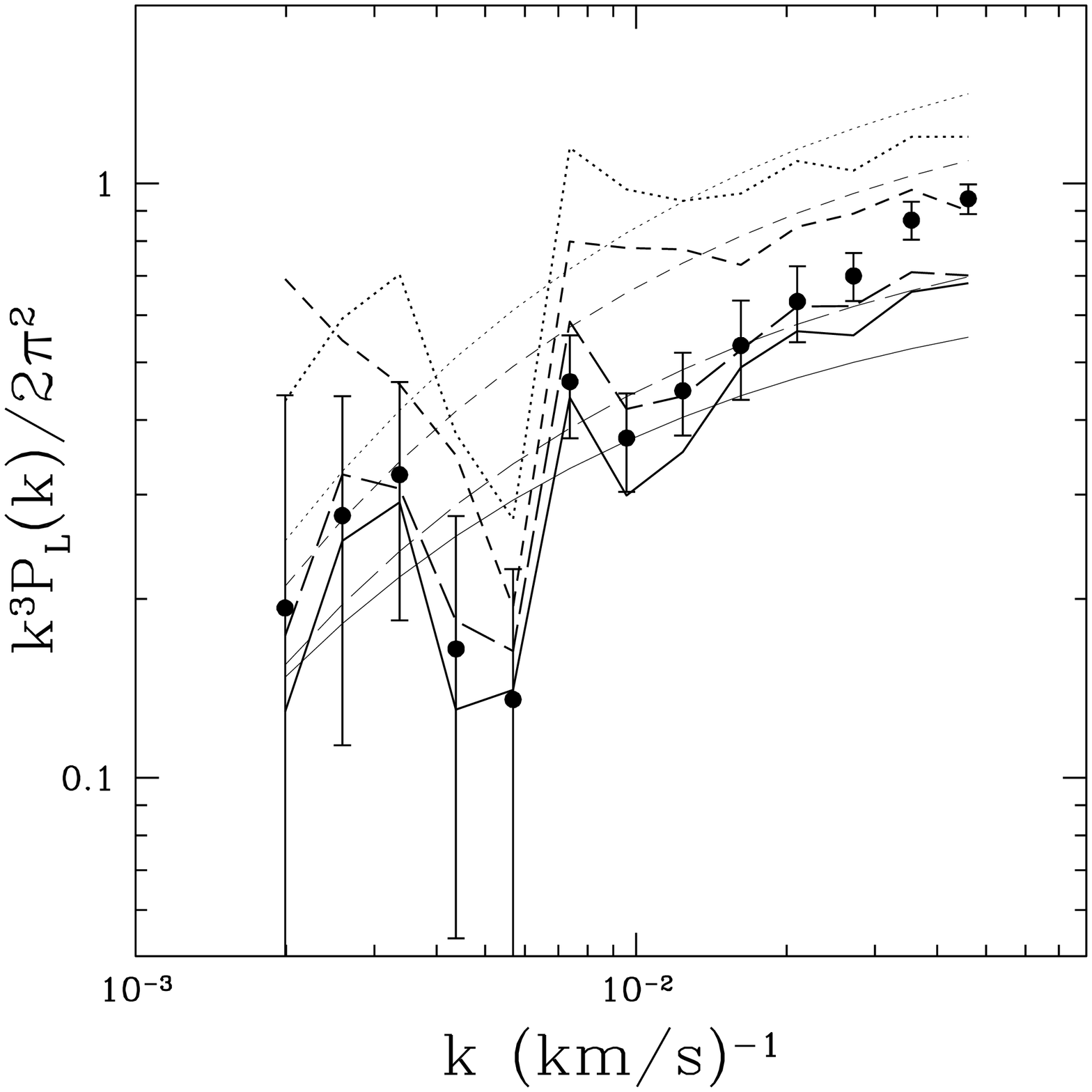}
\caption{\label{figDF}The flux power spectrum ({\it a\/}) and the
recovered linear power spectrum ({\it b\/}) in four 
different cosmological models:
Croft et al.\ \shortcite{CWB01} fiducial model ({\it solid line\/}),
an open model with $\Omega_{m,0}=0.45$ and $n=0.7$ ({\it long-dashed line\/}), 
an open model with $\Omega_{m,0}=0.2$ and $n=0.7$ ({\it long-dashed line\/}), 
and
an open model with $\Omega_{m,0}=0.1$ and $n=0.7$ ({\it dotted line\/}).
}
\end{figure}
Similarly, the dependence on the matter density parameter at z=2.72, 
$\Omega_{m,3}$,
was not tested by Croft et al.\ \shortcite{CWB01}.
This may not be an issue after all, because for any reasonable flat model with
a cosmological constant, $\Omega_{m,3}=1$ to better than 5\%. 
Nevertheless, we have tested the matter density dependence. 
We have run three open models
with $n=0.7$ and $\Omega_{m,0}=0.45$ ($\Omega_{m,3}=0.75$), 
$\Omega_{m,0}=0.2$ 
($\Omega_{m,3}=0.5$), and $\Omega_{m,0}=0.1$ ($\Omega_{m,3}=0.3$) respectively,
and the results of these simulations are presented in
Figure \ref{figDF}.
One can see that the recovered linear
power spectrum does depend on the assumed value of the density parameter
at $z=2.72$, even if all three models fit the observed flux power
spectrum.

The dependence on the density parameter cannot be rigorously
derived analytically,
but the form found empirically can be understood approximately on the basis
of arguments from linear theory.
The 
Gunn-Peterson optical depth $\tau$ \cite{GP65}
at every spatial point
is inversely proportional to the $du/dx$ along the line of sight
\cite{HGZ97}, where $u$ is the total velocity
(Hubble flow plus peculiar velocity) and $x$ 
is the comoving distance. In the linear regime
\begin{equation}
	{du\over dx} \sim {\partial\vec{u}\over\partial\vec{x}} =
	aH \left(1+f\delta\right),
\end{equation}
where $f \equiv d \ln D_+ / d \ln a
\approx\Omega^{0.6}$. The first term in the parenthesis is the
Hubble flow, while 
the second one is due to peculiar velocities. In the linear
regime the Hubble flow of course dominates, but at $z=3$ 
most of the 
Lyman-alpha forest is mildly nonlinear, with
$\Delta_k^2\equiv k^3P_L(k)/(2\pi^2) \sim 1$, as can be seen from
Fig.\ \ref{figDF}.

If we assume that different $k$ values are independent (which is
not quite true, but good enough for our hand-waving arguments),
we would expect that the power spectrum for the optical depth 
$\tau$ is approximately inversely
proportional to $(1+f\Delta_k)^2$. In other words, we hypothesize
that
the recovered linear power spectrum depends on cosmological
parameters in the following way:
\begin{equation}
	P_L^{\rm obs}(k) = P_L^{\rm fct}(k)
	\left(1+\Delta_k\over 
	1+f_3\Delta_k\right)^2,
	\label{denfac}
\end{equation}
where we will call $P_L^{\rm fct}(k)$
the ``factorized''
linear power spectrum, $f_3$ is the values of
$f$ at $z=2.72$, and
\begin{equation}
	\Delta_k\equiv \left(k^3P_L^{\rm obs}(k)\over 2\pi^2\right)^{1/2}
\end{equation}
and thus depends on $P_L^{\rm obs}(k)$. In this form $P_L^{\rm fct}(k)$
 should be nearly
independent of 
\na{
the value of the matter density parameter at $z=2.72$ $\Omega_{m,3}$.
}

\begin{figure}
\epsscale{1.0}
\plotone{\figdir/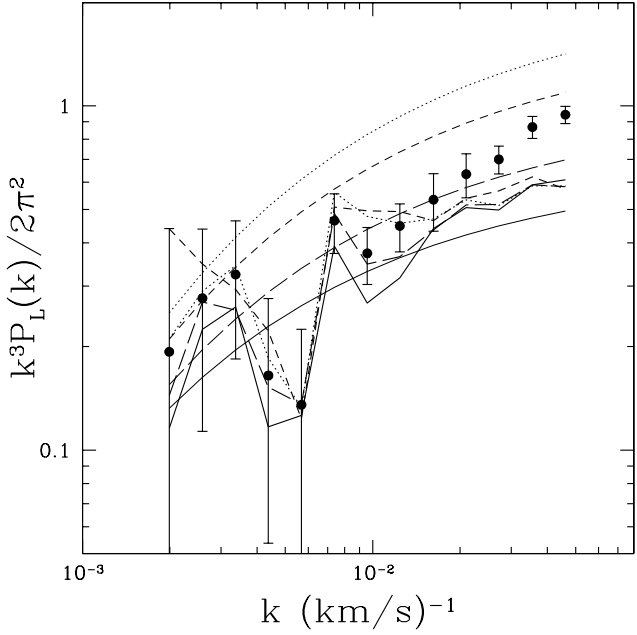}
\caption{\label{figAL}The factorized
recovered linear power spectrum in four cosmological models from Fig.\
\protect{\ref{figDF}}. Line markings are the same.
}
\end{figure}
Figure \ref{figAL} now gives $P_L^{\rm fct}(k)$ for the four models shown in
\ref{figDF}. As one can see, the factorization
(\ref{denfac}) does account for the main dependence of the recovered
linear power spectrum on cosmological parameters, although some differences
remain. It is however not
convenient to use in practice because it is nonlinear and implicit for
$P_L^{\rm obs}(k)$. But we can notice that for $\Omega_{m,3}$ substantially
different from 1, $\Delta_k$ is about unity for the range of scales where
the observational error-bars are small. It turns out that a comparable
factorization can be obtained if we adopt a fixed value $\Delta_k=1.4$ in 
equation (\ref{denfac}),
\begin{equation}
	P_L^{\rm obs}(k) = P_L^{\rm fct}(k)
	\left(2.4\over 
	1+1.4f_3\right)^2.
	\label{denfacsim}
\end{equation}
This fit is linear, so it is easier to use in 
joint parameter estimation, while it still recovers the density dependence
to about 2.2\% precision.

\subsubsection{The systematic error of the recovered linear power
spectrum}

\begin{figure}
\epsscale{1.0}
\plotone{\figdir/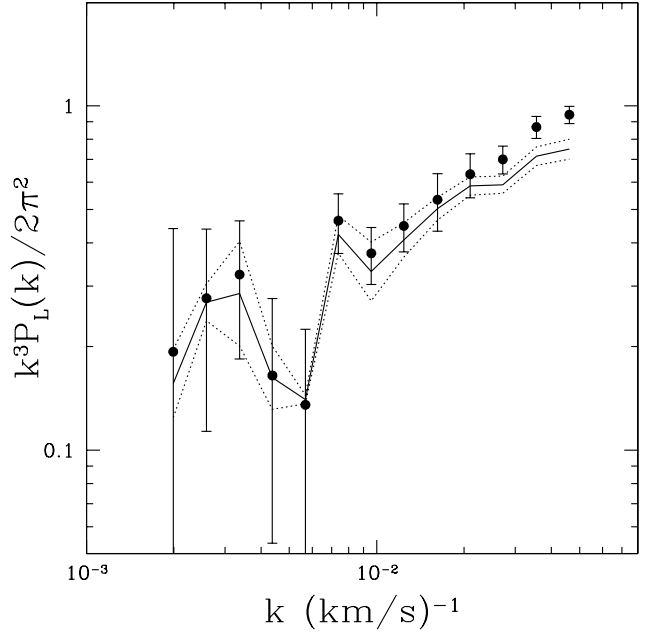}
\caption{\label{figAR}The ``best fit'' recovered linear power spectrum
({\it solid line\/}) and its 1 sigma systematic
error bars ({\it dotted lines\/}).
}
\end{figure}
We can now summarize our results. The recovered linear power spectrum
can be factorized in the following way:
\begin{equation}
	P_L^{\rm obs}(k) = P_L^{\rm fct}(k) Q_\Omega Q_T Q_\tau,
	\label{finfac}
\end{equation}
where
\begin{eqnarray}
	Q_\Omega & \approx & 	\left(2.4\over 
	1+1.4f_3\right)^2,\nonumber\\
	Q_T      & = & 20{,}000\dim{K}/T_0,\nonumber\\
	Q_\tau   & = & (0.349/\tau)^{0.75},
\end{eqnarray}
and $P_L^{\rm fct}(k)$ is independent of anything else (just numbers)
and is shown in Figure \ref{figAR} and in Table \ref{tabone}
together with its ``systematic'' error-bars.
We put the word ``systematic'' in quotes because
we considered only a subsample of all possible cosmological models, 
and our sampling of the parameter space is by no means uniform. We thus
suggest that errors quoted by us can be considered as an estimate
for the systematic error in the Croft et al.\ \shortcite{CWB01} measurement, but
the precise value of such an error and its covariance matrix may
depend very much on the specifics of the prior.

\begin{table}
\caption{The recovered factored linear power spectrum $P_L^{\rm fct}$
\label{tabone}}
\begin{tabular}{@{}cccc}
$\begin{array}{cc}k\\\dim{(km/s)}^{-1}\end{array}$ &
$\begin{array}{cc}P_L^{\rm fct}\\\dim{(km/s)}^3\end{array}$ &
$\begin{array}{cc}\mbox{Random}\\\mbox{err.}\dim{(km/s)}^3\end{array}$ &
$\begin{array}{cc}\mbox{Systematic}\\\mbox{err.}\dim{(km/s)}^3\end{array}$ \\
$1.99\times10^{-3}$ & $3.92\times10^8$ & $5.02\times10^8$ & $0.90\times10^8$ \\
$2.59\times10^{-3}$ & $3.06\times10^8$ & $1.80\times10^8$ & $0.38\times10^8$ \\
$3.37\times10^{-3}$ & $1.47\times10^8$ & $0.63\times10^8$ & $0.53\times10^8$ \\
$4.37\times10^{-3}$ & $3.85\times10^7$ & $2.59\times10^7$ & $0.82\times10^7$ \\
$5.68\times10^{-3}$ & $1.51\times10^7$ & $1.00\times10^7$ & $0.44\times10^7$ \\
$7.38\times10^{-3}$ & $2.08\times10^7$ & $0.41\times10^7$ & $0.26\times10^7$ \\
$9.58\times10^{-3}$ & $7.42\times10^6$ & $1.40\times10^6$ & $1.46\times10^6$ \\
$1.24\times10^{-2}$ & $4.23\times10^6$ & $0.67\times10^6$ & $0.49\times10^6$ \\
$1.62\times10^{-2}$ & $2.34\times10^6$ & $0.43\times10^6$ & $0.18\times10^6$ \\
$2.10\times10^{-2}$ & $1.25\times10^6$ & $0.18\times10^6$ & $0.08\times10^6$ \\
$2.72\times10^{-2}$ & $5.79\times10^5$ & $0.54\times10^5$ & $0.33\times10^5$ \\
$3.55\times10^{-2}$ & $3.15\times10^5$ & $0.23\times10^5$ & $0.20\times10^5$ \\
$4.61\times10^{-2}$ & $1.51\times10^5$ & $0.09\times10^5$ & $0.10\times10^5$ \\
\end{tabular}
\end{table}
The observational data \cite{RGS00,STR00,MMR01} sets the value of $Q_T$ at
\begin{equation}
	Q_T = 1 \pm 0.15
	\label{qtobs}
\end{equation}
and the error on $Q_T$ is uncorrelated with the systematic error on
$P_L^{\rm fct}(k)$.

With the mean optical depth the story is more complicated, because there is
no sufficiently accurate measurement of it. The PRS data give
$Q_{\tau,\rm PRS} = 1.00^{+0.11}_{-0.08}$, while the MCD data give
$Q_{\tau,\rm MCD} = 1.18\pm0.07$. Combining the two values in quadrature
(because the data sets they used are independent),
we find
\begin{equation}
	Q_\tau  = 1.11\pm 0.05.
	\label{qtauobs}
\end{equation}
And for the most conservative estimate, we can include the whole range of
quoted numbers:
\begin{equation}
	0.9 < Q_\tau < 1.25.
	\label{qtauobscons}
\end{equation}

\subsection{Band-Power Windows}
\label{bandpower}

\begin{figure}
\epsscale{1.0}
\plotone{\figdir/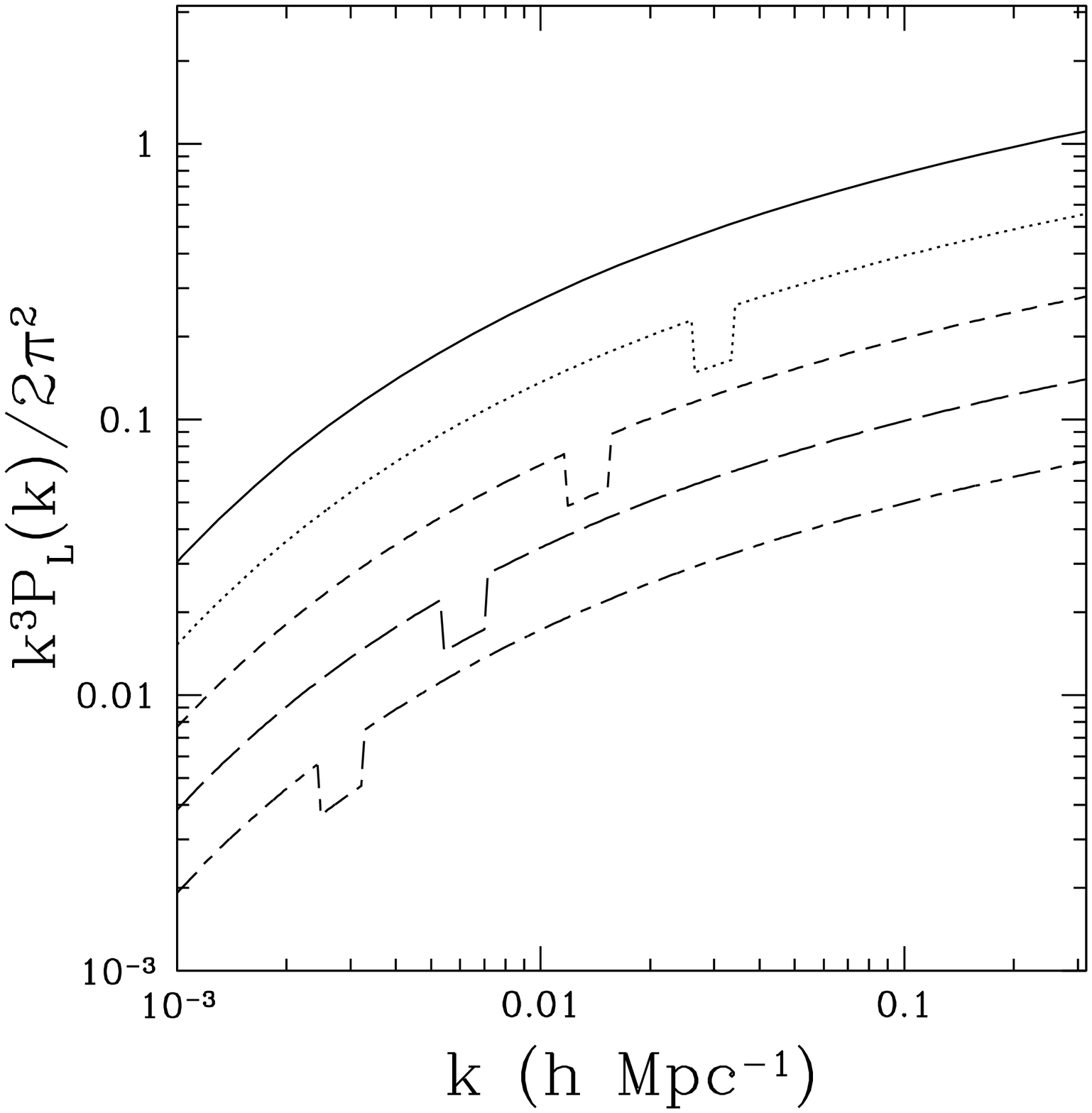}
\epsscale{1.0}
\plotone{\figdir/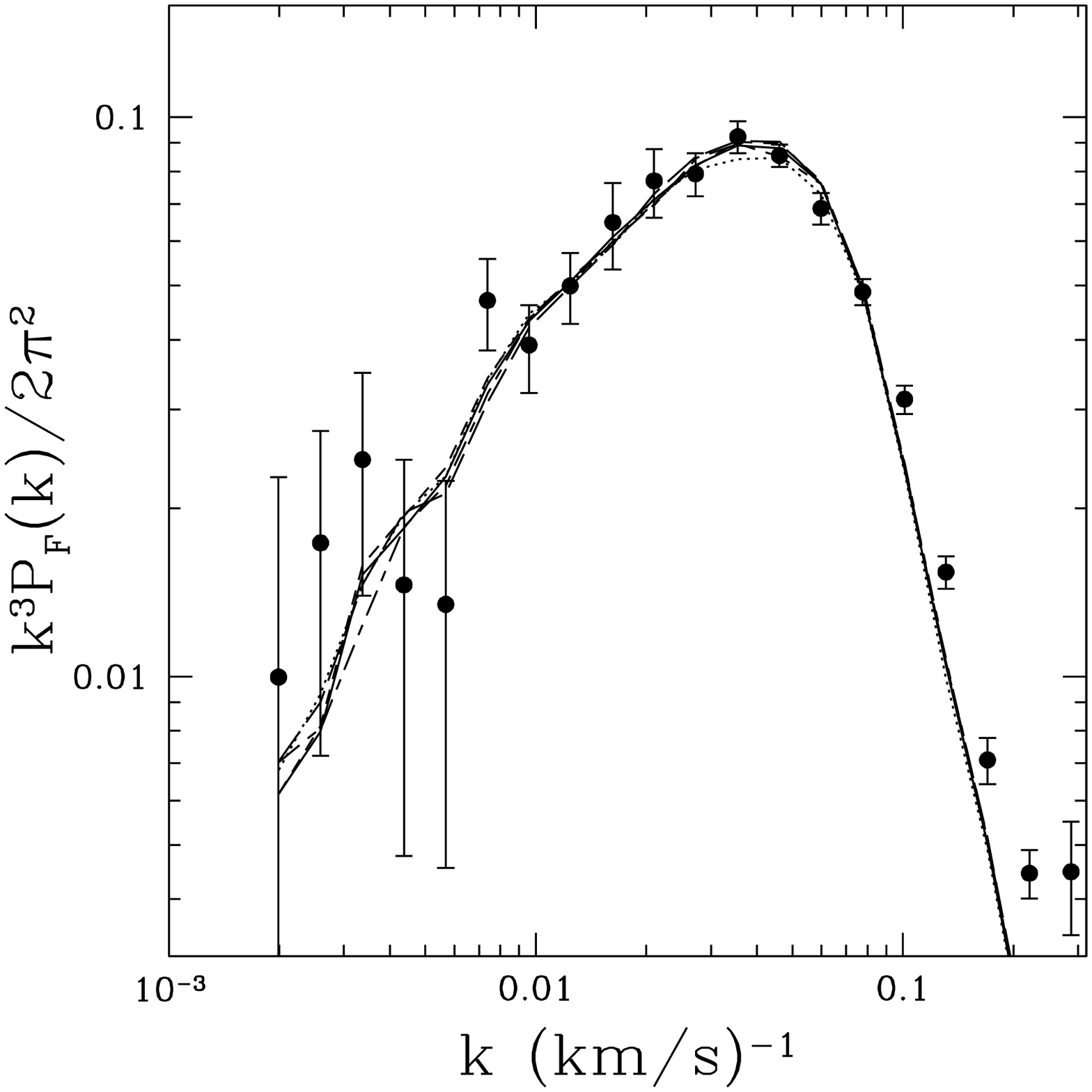}
\caption{\label{figCL}({\it a\/}) The assumed prior linear power spectra
for the fiducial Croft et al.\ model with the power reduced
at a particular value of the wavenumber. The power spectra are
offset vertically by 0.3 dex for clarity. ({\it b\/})
Their respective best-fit
flux power spectra.
}
\end{figure}
Measurements of flux power sample power
not at a single wavenumber, but rather over a finite band of wavenumbers.
The band power windows $b(k,k^\prime)^2$
in equation (\ref{pfll}) can be extracted by
differentiating the flux power spectrum $P_F(k)$ with respect to the
linear power spectrum $P_L(k)$.
The shape of the band-powers emerges most clearly if they are
scaled with the fiducial powers,
so we define scaled band-power windows by
\begin{equation}
	{\partial \ln P_F(k) \over \partial \ln P_L(k^\prime)}
	=
	b^2(k,k^\prime) P_L(k) / P_F(k^\prime).
	\label{dpfll}
\end{equation}

In order to measure the effective band-power windows in the present case,
we ran four additional models with the fiducial power spectrum of 
Croft et al.\ \shortcite{CWB01} reduced by 20\%
in four different wave-bands, corresponding to four $k$ values.
Linear power spectra for these models are shown in Figure \ref{figCL}a,
and their respective flux power spectra are shown in Fig.\ \ref{figCL}b.
The difference between the different models is comparable to the residual
uncertainty in the mean flux power spectrum (Fig.\ \ref{figRL}), and thus
we are only able to measure the difference between the various models
with the reduced power to an accuracy of about 30\%. A more
accurate determination of the 
band-power windows would require an implausibly large number of simulations.

As can be seen,
the abrupt reductions in the linear power spectrum
lead not to abrupt reductions in the flux power spectrum,
but rather to broad depressions.
\begin{figure}
\epsscale{1.0}
\plotone{\figdir/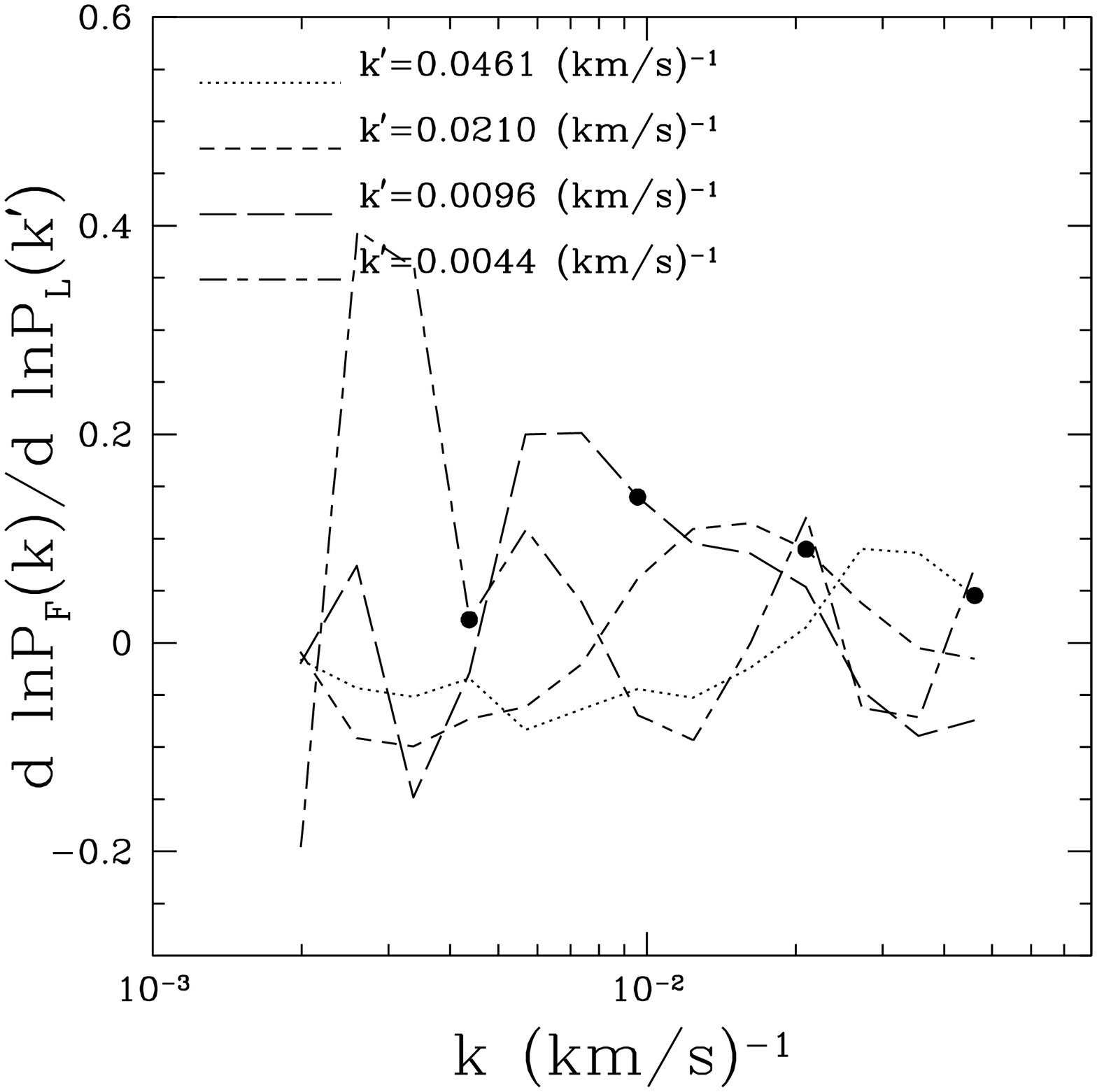}
\epsscale{1.0}
\plotone{\figdir/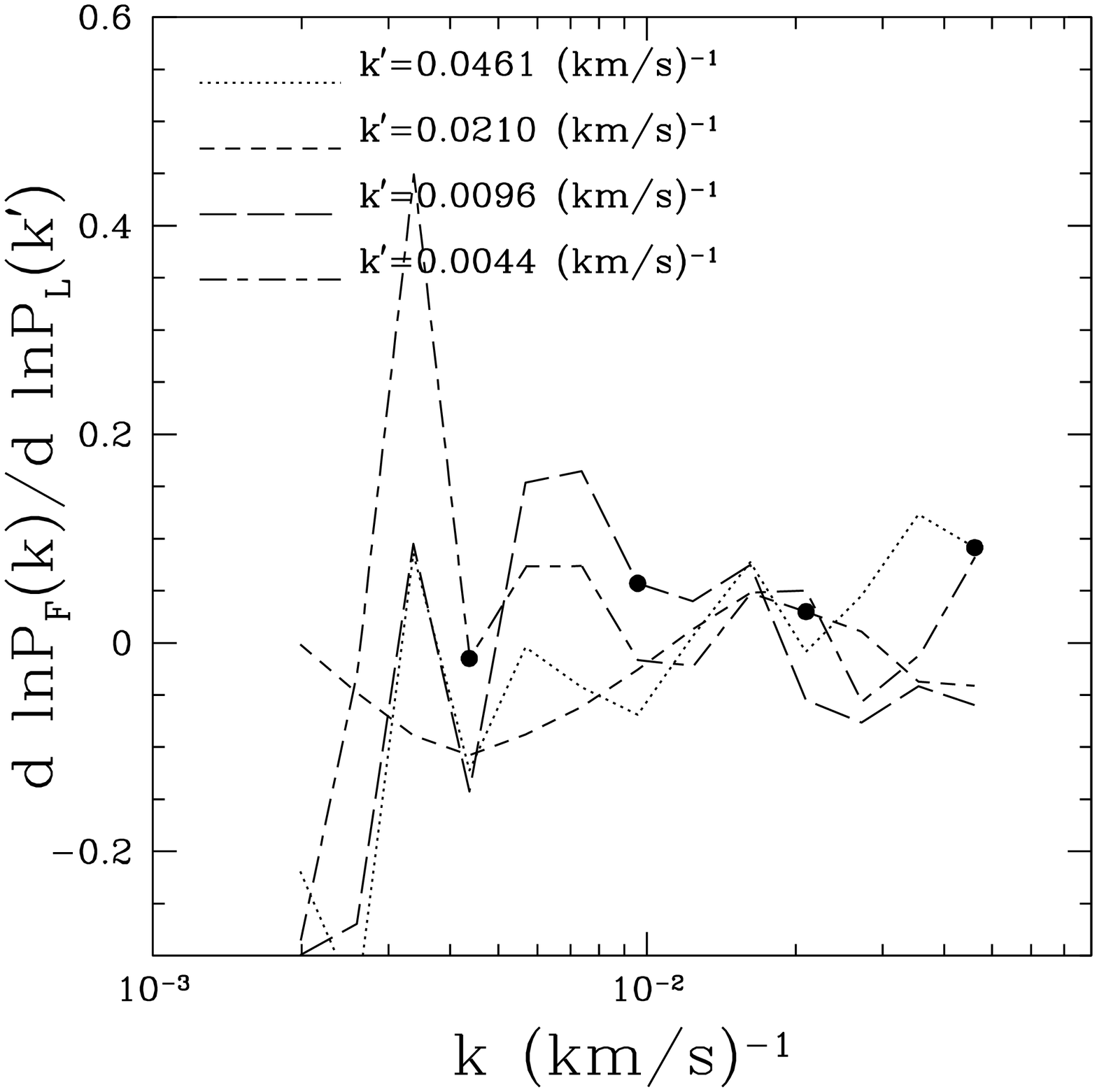}
\caption{\label{figCD}
The derivative of the measured flux power spectrum with respect to the
value of the input linear power spectrum at a given wavenumber
for four wavenumbers, as labeled.
The curves provide estimates of the effective band-power windows
of the flux power spectrum.
Filled circles show the values of $k^\prime$. Two cases are shown:
({\it a\/}) the amplitudes of the input power spectra are fixed to
$\sigma_8(z=2.72)=0.23$ and ({\it b\/}) the amplitudes of the
input linear power spectra
are adjusted to achieve the best fit to the observed flux power spectrum.
}
\end{figure}
Figure \ref{figCD} shows the inferred band-power windows themselves,
the derivative
of the flux power spectrum with respect to the input linear
power spectrum at a given wavenumber, both for the case when the amplitudes
of the input power spectra are kept fixed, and when the amplitudes are
adjusted to fit the observed flux power spectrum.
Evidently the effective band-power windows are quite broad.
\begin{figure}
\epsscale{1.0}
\plotone{\figdir/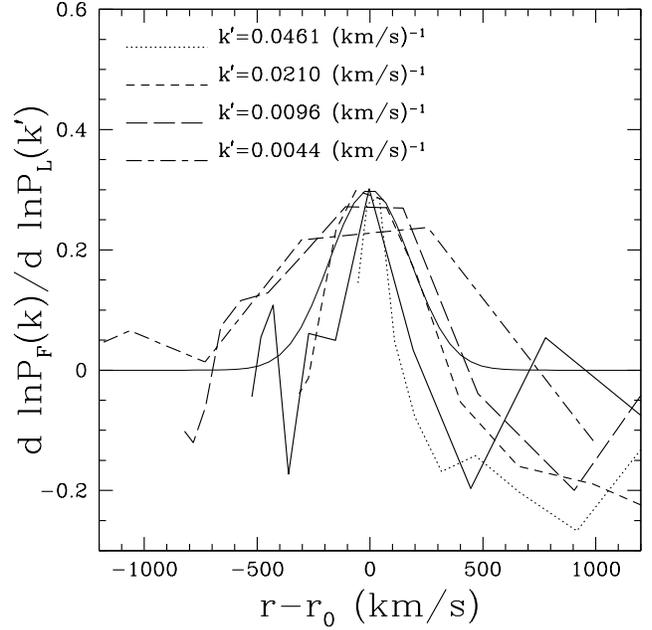}
\caption{\label{figCR}
The derivative of the measured flux power spectrum 
with respect to the
value of the input linear power spectrum at a given wavenumber
for four wavenumbers and for the fixed
the amplitude of the prior power spectra shown as a function of
$r-r_0$, where $r=2\pi/k$.
This is essentially Fig.\ \protect{\ref{figCD}} with a transformed $x$-axis.
The bold solid line shows what the long-dashed line becomes when
peculiar velocities are set to zero.
The curves are scaled vertically to have the same amplitude at maximum.
The thin solid line shows a gaussian with $\sigma=200\dim{km/s}$, which is
a fit to the short-dashed line.
}
\end{figure}

Fig.\ \ref{figCD} indicates that the band-power windows are broader
at smaller scales.
Figure \ref{figCR} attempts to quantify this feature by showing the
same derivatives as Fig.\ \ref{figCD}, except as a function of $r=2\pi/k$
rather than $k$.
To bring out the similarity of shapes,
the four band-power windows from Fig.\ \ref{figCD} have been shifted horizontally
so that their center points $r_0 = 2\pi/k_0$ coincide (the centers of the 
band-power windows $k_0$ do not necessarily fall on $k^\prime$,  
but are always within one bin in $k$ space, from which we conclude that
this discrepancy is most likely due to 
the finite $k$-space sampling),
and vertically to the same maximum.
The four band-power windows do not have exactly the same width in real
space. Fitting to a
Gaussian of width $\sigma$, we find that
$\sigma$ as a function of $k^\prime$ can be approximated as
\begin{equation}
	\sigma(k^\prime) \approx {25\dim{km/s}\over \sqrt{
	k^\prime\times1\dim{km/s}}}.
\end{equation}

In order to illustrate that it is indeed peculiar velocities that are
responsible for the smoothing of the recovered linear power spectrum,
we show in Fig.\ \ref{figCR} the derivative
of the flux power spectrum with respect to the prior linear
power spectrum at $k^\prime=0.0096\dim{(km/s)}^{-1}$ (an equivalent of the
long-dashed line) with all peculiar velocities
set to zero. In this case the peak is much narrower, and its width is
explained by finite sampling of the $k$-space alone. 

It is important to underscore here that it is the power spectrum itself
which is smoothed, and not the flux. Croft et al.\ \shortcite{CWB01} did consider the
effect of smoothing the Lyman-alpha absorption spectrum (Fig.\ 10 in their
paper), but the effect that we discuss here is different -- it is
a correlation of power
between the neighboring $k$ values rather than a reduction
in power on small scales.

This effect is easy to understand physically. Let us imagine that the
matter power spectrum is a delta function, i.e.\ the power is only
nonzero at a given value $k_0$. In the absence of peculiar velocities
the three-dimensional flux power spectrum would be non-zero only in a
narrow range of scales around $k_0$ (we cannot claim that this range is
infinitesimally small because the relationship between the linear matter
power spectrum and the Lyman-alpha flux power spectrum is nonlinear; however,
the solid line in Fig.\ \ref{figCR} does demonstrate that this range
is less than the $k$-space sampling in Croft et al.). In
the presence of peculiar velocities a given structure in physical space would
appear shifted along the line of sight in redshift space (which is where the 
flux is measured), thus smearing power over a range of scales comparable to 
the 1D velocity dispersion.

\subsection{Missing physics}

In the previous subsections we have considered only one possible source of
the systematic error, namely the choice of the prior linear power spectrum.
Another possible uncertainty comes from the fact that no simulation 
includes all of the relevant physics. We can identify at least three major
physical ingredients that are missed by pure PM simulations:
\begin{enumerate}
\item gas pressure;
\item inhomogeneities in the ionizing background;
\item shocks in the gas.
\end{enumerate}
We discuss these effects in that order.

\subsubsection{Gas pressure}

The role of gas pressure is well understood. In the linear
regime it can simply be expressed by the filtering scale of baryonic 
perturbations, $k_F$:
\begin{equation}
	P_{\rm gas}(k) \approx P_{\rm DM}(k)e^{-2k^2/k_F^2},
	\label{bfil}
\end{equation}
where $k_F$ is a complicated function of the entire thermal history
of the universe \cite{GH98}, 
and equation (\ref{bfil}) is a good approximation
to the exact solution to the linear theory equations for $k<0.8k_F$.
There are also strong reasons to believe that the filtering scale
describes the effect of gas pressure even in the nonlinear regime
\cite{G00b}.

Recent measurements of the thermal history of the universe
\cite{RGS00,STR00,MMR01}
give $k_F=(35\pm5)h\dim{Mpc}^{-1}$, significantly smaller than the range
of scales $k\la 3h\dim{Mpc}^{-1}$
measured from the Lyman-alpha forest power spectrum 
(the reason being that pressure filtering, always lagging behind the growth
of the Jeans mass, is subdominant to the thermal broadening of the
absorption spectrum). Thus, the pressure filtering produces only about
2\% correction to the recovered linear power spectrum, and it is unmeasurable
at the current time. This is, of course, in a full agreement with the 
conclusions of Croft et al.\ \shortcite{CWB01}.

\subsubsection{Inhomogeneities in the ionizing background}
\label{sec:inh}

Inhomogeneities in the ionizing background can come from two classes
of sources: large scale inhomogeneities from quasars, and small scale
inhomogeneities from galaxies. The former is known to be unimportant
\cite{CWP99}, but the effect of the latter can only be estimated from
advanced numerical simulations which include effects of the radiative
transfer. Here we use a simulation from Gnedin \shortcite{G00a}. 
We follow the
Croft et al.\ \shortcite{CWB01} procedure for the last output of this simulation at $z=4$,
but we analyze it two different ways. In the first case we
use the neutral hydrogen fraction 
from the simulation data (which reflects the non-uniformity of the ionizing
radiation background)
to compute a synthetic Lyman-alpha forest at $z=4$.
In the second case we assume that the ionizing radiation is uniform
and compute the neutral hydrogen fraction from ionization equilibrium.
The value of the background is calculated by normalizing to the mean
opacity in the simulation, so that in both cases the mean opacity is
the same.
In both cases we adopt the gas temperature as given in the simulation.
After computing the flux power spectra in both cases,
we find that they differ by less than 1\%. 

The Gnedin \shortcite{G00a} simulation suffers from its small box size 
($4h^{-1}\dim{Mpc}$), and therefore 
a 1\% difference is probably an underestimate,
but it appears unlikely that it underestimates the role of inhomogeneities
in the ionizing background by, say, a factor of 10. We therefore conclude
that inhomogeneities in the ionizing background do not bias the measurement
of the matter power spectrum from the Lyman-alpha forest.

Surprisingly, while the power spectra in the two calculations
are very similar, the actual value
of the volume averaged photoionization rate 
differs by 20\%, being larger for the inhomogeneous ionizing background.
Thus, simulations that assume homogeneous ionization and adjust
the photoionization rate to fit the mean opacity of the forest
actually underestimate the photoionization rate by at least 20\%,
or, equivalently,
overestimate the baryon density by at least 10\% (and probably by a
significantly larger factor). This may explain 
larger numbers for the baryon density typically 
found in such simulations \cite{HMK01}.

One of the possible concerns with this conclusion is that 
the Gnedin \shortcite{G00a} simulation was terminated at $z=4$, whereas the
Croft et al.\ \shortcite{CWB01} measure the linear matter power spectrum
at $z=3$. 

However, observationally we know that
the following three quantities do not change by more than a factor of two
between $z=4$ and $z=3$: the mean photoionization rate 
\cite{LSW96,CEC97,SBD00}, the star formation rate
\cite{SAC99}, and the mean separation between star forming
galaxies (deduced from constancy of the mean UV luminosity density and the
star formation rate). This immediately implies that
the proximity regions of star forming galaxies occupy approximately the
same fraction of the total volume at $z=4$ and $z=3$, and thus expected
fluctuations in the ionizing background are comparable at $z=4$ and $z=3$.
We thus can be confident that our conclusions about the role of
fluctuations in the ionizing background at $z=4$ also hold at $z=3$.

\subsubsection{Shocks and other mess}

Obviously, non-gravitational effects like shocks can significantly
affect the distribution of the Lyman-alpha forest. Fortunately,
we have observational data \cite{RGS00,STR00,MMR01}
that suggest that non-gravitational effects
are not significant in the forest. First, the IGM temperature at $z\sim3$
is quite low, only $20{,}000\dim{K}$, and is entirely consistent with
pure photoheating of the gas. Second, the narrowness of the cut-off
in the Doppler parameter vs column density distribution of the Lyman-alpha
lines suggests that the photoionization-induced ``effective equation of state''
has a small scatter, again fully consistent with the pure photoheating of 
the gas. Any realistic shock heating would destroy this tight correlation.

Thus, while we cannot rigorously prove that shocks and other 
non-gravitational effects can be neglected in modeling the Lyman-alpha
forest, they would require a rather special fine-tuning which closely
mimics the work of gravity and photoionization.

\section{Conclusions}

The two main drawbacks of the Croft et al.\ \shortcite{CWB01} paper
are (a) that it misses the dependence of the recovered linear
power spectrum on cosmological parameters
(eq.\ [\ref{finfac}]) and
(b) that it misses the fact that the effective band-power windows
of the flux power spectrum are broad.
The broad band-powers must induce broad correlations between estimates of
the flux power, and consequently also of the recovered matter power spectrum.
But as the whole, we must acknowledge
that the measurement that they present is remarkable -- it currently offers
the best measurement of the linear matter power spectrum 
over a range of scales that are nonlinear today.

The effective band-power windows can be approximated as Gaussians
with a dispersion $\Delta k = k^{3/2}/(2\pi)\times 25\dim{(km/s)}^{1/2}$ 
in Fourier space:
\begin{equation}
	P_L^{\rm REC}(k) = {1\over \Delta k\sqrt{2\pi}}\int_0^\infty
	 dk^\prime\,
	 P_L^{\rm TRUE}(k^\prime)\,e^{\displaystyle-{(k-k^\prime)^2\over
	2\Delta k^2}},
\end{equation}
where $P_L^{\rm REC}(k)$ is the recovered and
$P_L^{\rm TRUE}(k)$ is the true underlying linear power spectra respectively.
The smoothing is caused by peculiar velocities.

And while the resulting strong covariance between estimates of
power at different wavenumbers 
will make careful comparison of the data and cosmological models more
complicated than simple $\chi^2$ fitting, this measurement will 
undoubtedly play an important role in securing
accurate values of the cosmological parameters.

We also notice in passing that residual inhomogeneities in the ionizing
background do not affect the measurement of the linear power spectrum,
but may substantially bias estimates of the cosmic baryon density
from the Lyman-alpha forest data.

We thank Rupert Croft and David Weinberg for valuable comments. We are
also grateful to Rupert for providing the revised version of their paper
prior to posting it to astro-ph.
This work was partially supported by NASA ATP grant NAG5-10763 and by the
National Computational Science
Alliance under grant AST-960015N, and utilized the SGI/CRAY Origin 2000 array
at the National Center for Supercomputing Applications (NCSA).

\end{document}